\documentclass[preprint]{revtex4}
\usepackage{graphicx}
\usepackage{bm}
\usepackage{dcolumn}
\usepackage{amsmath}
\def \lleq {\lower0.9ex\hbox{ $\buildrel < \over \sim$} ~}
\def \ggeq {\lower0.9ex\hbox{ $\buildrel > \over \sim$} ~}

\def \beq  {\begin{equation}}
\def \eeq  {\end{equation}}
\def \ber  {\begin{eqnarray}}
\def \eer  {\end{eqnarray}}

\begin{document}
\newcommand{\newc}{\newcommand}

\newc{\be}{\begin{equation}}
\newc{\ee}{\end{equation}}
\newc{\ba}{\begin{eqnarray}}
\newc{\ea}{\end{eqnarray}}
\newc{\bea}{\begin{eqnarray*}}
\newc{\eea}{\end{eqnarray*}}
\newc{\D}{\partial}
\newc{\ie}{{\it i.e.} }
\newc{\eg}{{\it e.g.} }
\newc{\etc}{{\it etc.} }
\newc{\etal}{{\it et al.}}
\newcommand{\nn}{\nonumber}
\newc{\ra}{\rightarrow}
\newc{\lra}{\leftrightarrow}
\newc{\lsim}{\buildrel{<}\over{\sim}}
\newc{\gsim}{\buildrel{>}\over{\sim}}
\title{Figure of Merit and Different Combinations of Observational Data Sets}
\author{Qiping Su$^{1,2}$}
\email{sqp@hznu.edu.cn}
\author{Zhong-Liang Tuo$^{2}$}
\email{tuozhl@itp.ac.cn}
\author{Rong-Gen Cai$^{2}$}
\email{cairg@itp.ac.cn}
\affiliation{
$^{1}$  Department of Physics, Hangzhou Normal University, Hangzhou, 310036, China  \\
$^{2}$ Key Laboratory of Frontiers in Theoretical Physics, Institute of
Theoretical Physics, Chinese Academy of Sciences, P.O. Box 2735,
Beijing 100190, China }
\date{\today}

\begin{abstract}
To constrain cosmological parameters, one often makes a joint
analysis with different combinations of observational data sets. In
this paper we take the figure of merit (FoM) for Dark Energy Task
Force fiducial model (CPL model) to estimate goodness of different
combinations of data sets, which include 11 widely-used
observational data sets (Type Ia Supernovae, Observational Hubble
Parameter, Baryon Acoustic Oscillation, Cosmic Microwave Background,
X-ray Cluster Baryon Mass Fraction, and Gamma-Ray Bursts). We
analyze different combinations and make a comparison for two types
of combination based on two types of basic combinations, which are
often adopted in the literatures. We find two sets of combinations,
which have strong ability to constrain the dark energy parameters,
one has the largest FoM, the other contains less observational data
with a relative large FoM and a simple fitting procedure.
\end{abstract}

\pacs{98.80.Es, 95.36.+x, 98.80.-k}
\maketitle
%%%%%%%%%%%%%%%%%%%%%%%%%%%%%%%%%%%%%%%%%%%%%%%%%%%
%%%%%%%%%%%%%%%%%%%%%%%%%%%%%%%%%%%%%%%%%%%%%%%%%%%%%
\section{Introduction}
Since the discovery of cosmic acceleration expansion of the
universe~\cite{Riess:1998cb,perlmutter99}, its origin has been a hot topic in
modern cosmology and theoretical physics.  Various dark energy (DE)
models explaining the acceleration expansion have been
proposed~\cite{Li:2011sd,Copeland:2006wr,Caldwell:1997mh,Steinhardt:1999nw,Capozziello:2003tk,Li:2004rb,a1,a2,a3,a4,a5,a6},
which originate from extra dimensions, string theory, supergravity,
extended standard model of particle physics, and so on. Most of the
DE models can fit astronomical observational data.  At present, the
main task of DE studies is to constrain the present equation of
state (EoS) of DE $w_{de0}$, the fractional energy density
$\Omega_{de0}$ and to give the possible evolution form of
$w_{de}(z)$. Several parameterization forms of $w_{de}(z)$ have been
proposed, such as the CPL parameterization
$w_{de}(z)=w_0+w_1z/(1+z)$~\cite{Chevallier:2000qy,Linder:2002et}
and the parameterization $w_{de}(z)=w_0+w_1z$~\cite{Huterer:2000mj,b1,b2,b3}.
Fitting the parameterization models of DE with present observational
data, constraints of DE can be obtained. The fitting results favor
that the present value of $w_{de}$ is very close to $-1$ and its
variation is very slow. But it still not be able to confirm whether
$w_{de}$ is evolving or just a constant. Of course, these
constraints are also dependent on the parameterization forms used.
As a result several methods have been proposed to get
model-independent constraints of DE from observations
~\cite{Huterer:2002hy,Huterer:2004ch,Hojjati:2009ab,Wang:2009sn,Holsclaw:2010sk,c1,c2},
such as the uncorrelated bandpower estimate
(UBE)~\cite{Sullivan:2007pd} and the spline
method~\cite{Cai:2010qp,d1,d2}.

On the other hand, the fitting results also depend on the
observational data used and there might exist ``tensions'' among the
estimated EoS parameters from different data
sets~\cite{Sanchez:2009ka,e1,e2,e3,e4,e5,e6}.
 A single observation still gives poor constraints on the parameters of cosmological models.
For example, Type Ia supernovae observation, which has large number
of datapoints, gives poor constraints on $\Omega_{m0}$ and $w_{de}$
because of the degeneracy between $\Omega_{m0}$ and
$w_{de}$\cite{Maor:2000jy}. To break the degeneracy and get good
constraints of the parameters, one always combines different types
of observation to fit cosmological models. It is not surprising that
fitting a DE model with different combinations of observational data
sets, one would get different constraints on DE. There seemingly
does not exist a general rule to combine different observations for
fitting the cosmological parameters. One uses different combinations
in different works.

In this paper we would like to make a comparison for different
combinations of widely-used data sets for DE, which contain data of
Type Ia Supernovae (SN), Observational Hubble Parameter (HUB),
Baryon Acoustic Oscillation (BAO), Cosmic Microwave Background
(CMB), X-ray Cluster Baryon Mass Fraction (CBMF), Gamma-Ray Bursts
(GRB), etc. The fitting properties of observational data will be
studied by using the Dark Energy Task Force
(DETF)~\cite{Albrecht:2006um,f1,f2} fiducial model (CPL model), i.e., all
combinations of data will be fitted with the CPL parameterization.
The best-fitted values of $\Omega_{m0}, w_0, w_1$ and their 68\%,
95\% C.L. errors for each combination will be given. To estimate
constraints on $w_{de}$ and compare the goodness of data, we will
calculate the figure of merit (FoM)~\cite{Albrecht:2006um,f1,f2} for each
combination, which is proportional to the inverse area of the error
ellipse in the $w_0\sim w_1$ plane. Namely we will take the FoM as
the diagnostic to quantify the goodness of each combination of
observational data to constrain the DE parameters.

The paper is organized as follows. In Section II, we introduce 11
sets of observational data and study their fitting properties
 on the basis of Type Ia supernovae Union2 data~\cite{Amanullah:2010vv}.
In Section III we  study the effect of different combinations of
data sets on constraining DE parameters. We divide the combinations
into two types, based on two basic combinations, SN+BAOI+CMB and
SN+A+R, respectively. The properties of various combinations will be
analyzed and compared, and the best combination (which has the
largest FoM) will be given.  Main results are summarized in  Section
IV.

\section{Observational Data Sets}

One of important astronomical observations  to constrain DE is the
distance measurement (such as SN and shift parameter R). To constrain DE models with these
data, one must give the form of Hubble function $H(z)$ (or
$E(z)=H(z)/H_0$). In a flat Friedmann-Robertson-Walker (FRW)
universe, one has
 \be
E^2(z)=\Omega_{r0}(1+z)^4+\Omega_{b0}(1+z)^3+\Omega_{dm0}
(1+z)^3+\Omega_{de0}F(z),
  \ee where $\Omega_{r0}$,
$\Omega_{b0}$, $\Omega_{dm0}$ and $\Omega_{de0}$ are present values
of the fractional energy density for radiations, baryons, dark
matter and dark energy respectively, with
$\Omega_{r0}+\Omega_{b0}+\Omega_{dm0}+\Omega_{de0}=1$. The energy
densities of baryons and dark matter are often written together as
$\Omega_{b0}+\Omega_{dm0}=\Omega_{m0}$. The radiation density is the
sum of photons and relativistic neutrinos~\cite{Komatsu:2008hk}:
\be
\Omega_r^{(0)}=\Omega_{\gamma}^{(0)}(1+0.2271N_{n}),
\ee
where $N_{n}$ is the number of neutrino species and
$\Omega_{\gamma}^{(0)}=2.469\times10^{-5}h^{-2}$ for
$T_{cmb}=2.725K$ ($h=H_0/100~Mpc \cdot km\cdot s^{-1}$). The
evolving function $F(z)$ for DE depends on the equation of state,
$w_{de}(z)$, of DE models,
\be
F(z)=e^{3\int_0^z\frac{1+w_{de}}{1+x}dx}. \ee We will fit
observational data with the CPL parameterization. In that case, one
has
\be
F(z)=(1+z)^{3(1+w_0+w_1)}\exp[-\frac{3w_1z}{1+z}].
\ee
When fitting the CPL model with data of distance measurement, the
parameters $\Omega_{m0}$ (or $\Omega_{de0}$),$w_0$, $w_1$, and $h$
are needed. While for some other data, such as those of CMB and BAO,
one needs to calculate physical quantities other than the distance
(such as the redshift of decoupling, $z_s$, and the comoving sound
horizon at decoupling, $r_s$) and the parameter $\Omega_{b0}$ must
be involved. Whenever the fitting procedure needs to use
$\Omega_{b0}$, one more datapoint will always be adopted to
constrain $\Omega_{b0}$ and $h$~\cite{Burles:2000zk}:
\be
\Omega_{b0}h^2=0.022\pm0.002~.
\ee
In addition, in all calculations we will assume the prior that the
age of the universe $T_0$ satisfies $10Gyr<T_0<20Gyr$.

We will use Union2 SN data~\cite{Amanullah:2010vv} and
other widely-used 10 observational data sets to constrain the CPL
model.  When studying the fitting properties of these 10
observational data, we will combine each data set with the SN data
to fit CPL model. Since we are only interested in the constraints on
DE, only the best-fitted values of $\Omega_{m0}$,$w_0$, $w_1$ and
their $68\%$, $95\%$ C.L. errors will be given, which are obtained
by using the Markov Chain Monte Carlo method. To estimate and
compare the goodness of constraints from the observational data
sets, we calculate the  figure of merit
(FoM)~\cite{Albrecht:2006um,f1,f2,Wang:2008zh} for each combination of
data sets, which is proportional to the inverse area of the error
ellipse in the $w_0\sim w_1$ plane:
\be
{\rm FoM}=\left[{\rm det}C(w_0,w_1)\right]^{-1/2}~,
\ee
where $C(w_0,w_1)$ is the covariance matrix of $w_0$ and $w_1$ after
marginalizing out all of the other cosmological parameters. Larger
FoM means stronger constraint on the parameters since it corresponds
to a smaller error ellipse.

%%%%%%%%%%%%%%%%%%%%%%%%%%%%%%%%%%%%%%%%%%%%%%%%%%%%%%%%%%%%%%%%%%%%%%%%%%%%%%%%%%%%%%%%%%%
\begin{table}
\begin{centering}
\begin{tabular}{|c|c|c|c|c|c|}
\hline
data & $\Omega_{m0}$ & $w_0$ & $w_1$ & $\chi^2_{min}$ & FoM\\
\hline
\hline
SN &$0.419_{-0.028-0.238}^{+0.090+0.133}$&$-0.86_{-0.32-0.59}^{+0.46+1.19}$&
$-5.51_{-8.79-24.09}^{+6.99+7.59}$&541.431&0.647\\
\hline
SN+GRB&$0.413_{-0.035-0.310}^{+0.091+0.132}$&$-0.88_{-0.30-0.54}^{+0.42+1.16}$&$-5.02_{-7.52-19.86}^{+6.48+7.13}$&564.727&0.700\\
\hline
SN+CBMF&$0.378_{-0.025-0.097}^{+0.085+0.126}$&$-0.97_{-0.23-0.47}^{+0.31+0.71}$&$-2.77_{-4.90-13.09}^{+2.93+4.36}$&583.725&1.356\\
\hline
SN+HUB&$0.281_{-0.027-0.217}^{+0.134+0.170}$&$-1.00_{-0.12-0.34}^{+0.22+0.47}$&$-0.21_{-2.82-8.46}^{+1.54+1.98}$&550.391&2.315\\
\hline
SN+CMB&$0.274_{-0.017-0.036}^{+0.025+0.051}$&$-1.01_{-0.14-0.34}^{+0.16+0.29}$&$0.01_{-0.84-1.54}^{+0.73+1.45}$&542.697&21.688\\
\hline
SN+R&$0.277_{-0.021-0.040}^{+0.021+0.045}$&$-1.01_{-0.14-0.30}^{+0.17+0.31}$&$-0.06_{-0.84-1.64}^{+0.75+1.38}$&542.642&21.376\\
\hline
\end{tabular}
\par\end{centering}

\caption{\label{tab1}The best-fitted values with 68\% and 95\% C.L. errors of $\Omega_{m0},\,w_0,\,w_1$,
and figure of merit for different combinations of data sets in the CPL model.} \label{I}
\end{table}

\subsection{Type Ia Supernovae (SN)}

 Type Ia supernovae observation gives the direct evidence for the existence of the acceleration expansion of
 the universe. In this work we take the Union2 SN dataset~\cite{Amanullah:2010vv}, for which $\chi_{SN}^{2}$ is given by
\be
\chi_{SN}^{2}=\sum_{i=1}^{557}\frac{[\mu_{th}(z_{i})-\mu_{ob}(z_{i})]^{2}}{\sigma^{2}(z_{i})}~,\label{musn}
\ee
where the theoretical distance modulus $\mu_{th}$ is defined as
\be
\mu_{th}(z)=5\log_{10}D_L+\mu_{0}~,~~\mu_{0}=42.384-5\log_{10}h~,\nonumber\\
\ee where the luminosity distance $D_L=(1+z)\int_{0}^{z}dx/E(x)$.
One can expand Eq.(\ref{musn}) with respect to $\mu_0$ as
\be
\chi_{SN}^{2}=\tilde A+2\tilde B\mu_{0}+\tilde C\mu_{0}^{2}
\ee
where
\begin{eqnarray}
\tilde A & = & \sum_{i=1}^{557}\frac{[\mu_{th}(z_{i};\mu_{0}=0)-\mu_{ob}(z_{i})]^{2}}{\sigma^{2}(z_{i})},\nonumber\\
\tilde B & = & \sum_{i=1}^{557}\frac{\mu_{th}(z_{i};\mu_{0}=0)-\mu_{ob}(z_{i})}{\sigma^{2}(z_{i})},\label{abc}\\
\tilde C & = & \sum_{i=1}^{557}\frac{1}{\sigma^{2}(z_{i})}~.\nonumber
\end{eqnarray}
The $\chi_{SN}^{2}$ has a minimum as \be
\tilde{\chi}_{SN}^{2}=\tilde A-{\tilde B}^2/{\tilde C}~.\label{sn}
 \ee
  We will use
$\tilde{\chi}_{SN}^{2}$ instead of $\chi_{SN}^{2}$, this way the
nuisance parameter $\mu_0$ is reduced. It is equivalent to
performing an uniform marginalization over $\mu_0$.

In Table~\ref{I}, it shows that the SN data give a largish
$\Omega_{m0}$ and the value of FoM is rather small, which means that
the constraint on the parameters is weak. This is  mainly due to the
degeneracy between $\Omega_{m0}$ and $w_{de}$. Thus the SN data are
often combined with other types of observations, especially that can
constrain $\Omega_{m0}$ tightly, such as BAO and CMB.

\subsection{Gamma-Ray Bursts (GRB)}
Gamma-Ray Bursts (GRB) are the most intense explosions we have
observed in the universe, most GRB observed are in the range
$0.1<z<8.1$. Thus, GRB are considered to be a complementary probe
to SN and a hopeful probe at high redshift. Here we adopt the 59
Hymnium GRB at $z > 1.4$ from~\cite{Wei:2010wu}, which are obtained
by calibrating the 109 GRB with Amati relation, using the
cosmological independent method proposed in~\cite{Liang:2008kx}.

The operation of GRB datapoints is the same as that of SN. To fit
those data with DE models it should be combined with SN data, i.e.,
$\chi^2_{GRB}$ should be calculated together with $\chi^2_{SN}$ as
\be \tilde{\chi}_{SN+GRB}^{2}=\bar{A}-\bar{B}^2/\bar{C}~. \ee where
\begin{eqnarray}
\bar{A} & = & \sum_{j=1}^{557+59}\frac{[\mu_{th}(z_{j};\mu_{0}=0)-\mu_{ob}(z_{j})]^{2}}{\sigma^{2}(z_{j})},\nonumber\\
\bar{B} & = & \sum_{j=1}^{557+59}\frac{\mu_{th}(z_{j};\mu_{0}=0)-\mu_{ob}(z_{j})}{\sigma^{2}(z_{j})},\label{abcd}\\
\bar{C} & = & \sum_{j=1}^{557+59}\frac{1}{\sigma^{2}(z_{j})}~.\nonumber
\end{eqnarray}
The datapoints in the summation contain 557 SN and 59 GRB.

The results for SN+GRB are shown in Table I, which are almost the
same as that of SN. With addition of 59 GRB data, the increase of
$\chi^2$ (about 23) is much less than the increase of number of
datapoints, which implies that the present precision of the GRB
observation is not small enough.  As a result it means
that adding GRB data to SN data is not quite helpful to constrain
the DE model.

\subsection{X-ray Cluster Baryon Mass Fraction (CBMF)}

 The baryon mass fraction in clusters of
galaxies (CBMF) are also used to constrain cosmological parameters.
Here  we adopt the updated 42 observational $f_{gas}$ data
in~\cite{Allen:2007ue}. The X-ray gas mass fraction is defined as
the ratio of the X-ray gas mass to the total mass of a cluster,
which should be a redshift-independent constant. With the
$\Lambda$CDM reference cosmological model, one has \be
f_{gas}(z)=\frac{K\gamma
b}{1+s}\left(\frac{\Omega_b}{\Omega_m}\right)A\left[\frac{D_A^{\Lambda
CDM}(z)}{D_A(z)}\right]^{3/2}~,\label{fgas} \ee where $A$ is the
angular correction factor, which is due to the change in angle
subtended by $r_{2500}$ as the underlying cosmological model is
changed:
\be
A=\left(\frac{\theta_{2500}^{\Lambda CDM}}{\theta_{2500}}\right)^\eta\approx\left(\frac{H(z)D_A(z)}{[H(z)D_A(z)]^{\Lambda CDM}}\right)^\eta~.
\ee
Here we adopt the best-fitted average value $\eta=0.214$, and the
proper angular diameter distance is given by
\be
D_A(z)=\frac{D_L(z)}{(1+z)^2}~.
\ee
For other parameters K,$\gamma, b, s$ in Eq.(\ref{fgas}), please
refer to Refs.~~\cite{Allen:2007ue,Xu:2010zzb}. Here we introduce a
new parameter~$\lambda$~\cite{Nesseris:2006er}:
\be
\lambda\equiv\frac{K\gamma b}{1+s}(\frac{\Omega_b}{\Omega_m})~,
\ee
then one has
\be
f_{gas}(z_i)=\lambda A\left[\frac{D_A^{\Lambda CDM}(z_i)}{D_A(z_i)}\right]^{3/2}\equiv\lambda\widehat{f}_{gas}(z_i)~.
\ee
Next one can expand
\be
\chi^2_{CBMF}\equiv\sum_i\frac{(f_{gas}(z_i)-f_{gas,i}^{obs})^2}{\sigma^2_{i}}
\ee
with respect to $\lambda$ and obtain
\be
\chi^2_{CBMF}=\widehat{A}\lambda^2-2\widehat{B}\lambda+\widehat{C},
\ee
where 
\ba
\widehat{A}&=&\sum_i\frac{\widehat{f}_{gas}(z_i)^2}{\sigma^2_i},\nonumber\\
\widehat{B}&=&\sum_i\frac{\widehat{f}_{gas}(z_i) f_{gas,i}^{obs}}{\sigma^2_i},\\
\widehat{C}&=&\sum_i\frac{(f_{gas,i}^{obs})^2}{\sigma^2_i}~.\nonumber
\ea
 As in the case of SN, instead of $\chi^2_{CBMF}$, here we adopt the minimum of $\chi^2_{CBMF}$ with
respect to $\lambda$,
\be
\widehat{\chi}^2_{CBMF}=\widehat{C}-\widehat{B}^2/\widehat{A}~,
\ee
as $\chi^2$ for CBMF.

As shown in Table I,  the constraint from SN+CBMF is also weak, but
better than that of SN+GRB. The best fitted value of $\Omega_{m0}$
is still largish. The addition of 42 CBMF to SN data does not
improve much.

\subsection{Observational Hubble Data (HUB)}
The observational Hubble data can be obtained by using the
differential ages of passively-evolving galaxies as
\be
H\simeq-\frac{1}{1+z}\frac{\Delta z}{\Delta t}~.
\ee
We  use 12 observational Hubble data from~\cite{Stern:2009ep} and
\cite{Riess:2009pu}, which is tabled in~\cite{Xu:2010zzb}. The
chi-square is defined as:
\be
\chi^2_{HUB}=\sum_{i=1}^{12}\frac{[H_{th}(z_i)-H_{ob}(z_i)]^2}{\sigma_i^2}~.
\ee
In this case we use these values of $H(z_i)$ directly,  which can
break  the degeneracy between $\Omega_{m0}$ and $w_{de}$. With these
data,  as shown in Table I,  there is an increase of FoM and the
best-fitted $\Omega_{m0}$ becomes much smaller than in the case of
the SN data used only.

\subsection{Cosmic Microwave Background (CMB)}

1. CMB

In the CMB measurement, the distance to the decoupling epoch can be
accurately determined from the locations of peaks and troughs of
acoustic oscillations. For simplicity,  we use three parameters
$l_a,\ R,\ z_s$ obtained from WMAP7 (rather than the full data of
WMAP)~\cite{Komatsu:2010fb} to constrain cosmological models. Here
$z_s$ is the redshift of decoupling~\cite{Hu:1995en}, the shift
parameter $R$ is the scaled distance to the decoupling epoch:
  \be
R=\sqrt{\Omega_{m0}}\int_0^{z_s}\frac{dz}{E(z)},\label{r}
  \ee and $l_a$
is the angular scale of the sound horizon at the decoupling epoch:
  \be
\l_a=\pi\frac{r(a_s)}{r_s(a_s)},\label{la}
 \ee where $r(z)=\int_0^zdx/H(x)$
is the comoving distance and $r_s(a_s)$ is the comoving sound
horizon at the decoupling epoch:
  \be
r_s(a_s)=\int_0^{a_s}\frac{c_s(a)}{a^2H(a)}da,~~a_s=\frac{1}{1+z_s},\label{rs}
 \ee
  where the sound
speed $c_s(a)=1/\sqrt{3(1+\overline{R}_ba)}$ and
$\overline{R}_b=3\Omega_{b0}/4\Omega_{\gamma0}$ is the
photon-baryon energy density ratio.

The $\chi^2$ of the CMB data is constructed as: \be
\chi^2_{CMB}=X^TC_{M}^{-1}X \ee where \ba {X} &=&
\left(\begin{array}{c}
l_a - 302.09 \\
R - 1.725\\
z_s - 1091.3\end{array}
  \right)
  \ea
and the inverse covariance matrix
\begin{eqnarray}
 { C_{M}^{-1}}=\left(
\begin{array}{ccc}
2.305& 29.698& -1.333 \\
29.698& 6825.270& -113.180 \\
-1.333& -113.180& 3.414
\end{array}
\right)~.
\end{eqnarray}
In this case the parameter $\Omega_{b0}$ is involved. From the
definition of the shift parameter $R$ [see (\ref{r})], it is obvious
that $\Omega_{m0}$ can be well constrained, and the degeneracy
between $\Omega_{m0}$ and $w_{de}$ could be broken in some sense. As
shown in Table \ref{I}, the FoM and constraints of parameters have a
great improvement from those in the case of SN data used only.
Therefore the CMB data are a good supplementary of SN data in
constraining DE models.

2. Shift Parameter $R$

As an alternative, one can only adopt the shift parameter $R$ from
WMAP7~\cite{Komatsu:2010fb} to constrain the cosmological
parameters. In this case the corresponding $\chi^2$ is defined as
\be
\chi^2_{R}=\left(\frac{R-1.725}{0.018}\right)^2~.
\ee

It can be seen from Table~\ref{I} that the fitting results of SN+R
are very close to those of SN+R+$l_a+z_s$ (i.e., SN+CMB). Note that
 in this case one needs not to handle with $\Omega_{b0}$ and the
calculation is much simpler. The addition of a single datapoint $R$
to SN data greatly improves the fitting results, because $R$ can
alleviate the degeneracy between $\Omega_{m0}$ and $w_{de}$ well.

%%%%%%%%%%%%%%%%%%%%%%%%%%%%%%%%%%%%%%%%%%%%%%%%%%%%%%%%%%%%%%%%%%%%%%%%%%%%%%%%

\begin{table}
\begin{centering}
\begin{tabular}{|c|c|c|c|c|c|}
\hline
data & $\Omega_{m0}$ & $w_0$ & $w_1$ & $\chi^2_{min}$ & FoM\\
\hline
\hline
SN+BAOI&$0.420_{-0.066-0.177}^{+0.036+0.068}$&$-0.84_{-0.32-0.53}^{+0.23+0.71}$&$-5.79_{-2.60-10.25}^{+6.65+7.06}$&542.186&1.471\\
\hline
SN+BAOII&$0.428_{-0.033-0.359}^{+0.083+0.122}$&$-0.80_{-0.35-0.59}^{+0.46+1.20}$&$-6.50_{-8.78-26.82}^{+7.77+8.45}$&542.643&0.632\\
\hline
SN+BAOIII&$0.421_{-0.066-0.190}^{+0.035+0.066}$&$-0.82_{-0.33-0.55}^{+0.23+0.67}$&$-5.89_{-2.14-8.69}^{+6.73+7.13}$&542.134&1.514\\
\hline
SN+A &$0.278_{-0.018-0.037}^{+0.025+0.047}$&$-1.01_{-0.13-0.28}^{+0.17+0.33}$&
$-0.10_{-1.27-2.57}^{+1.04+1.89}$&542.643&13.223\\
\hline
SN+RBAO &$0.394_{-0.091-0.241}^{+0.026+0.056}$&$-0.93_{-0.21-0.41}^{+0.18+0.48}$&
$-3.78_{-0.038-4.35}^{+4.75+5.05}$&541.602&2.586\\
\hline
\end{tabular}
\par\end{centering}
\caption{\label{tab2}The best-fitted values with 68\% and 95\% C.L. errors of $\Omega_{m0},\,w_0,\,w_1$,
and figure of merit for different combinations of data sets in the CPL model.}
\label{II}
\end{table}

\subsection{Baryon Acoustic Oscillation (BAO)}

As baryons and photons are tightly coupled at early times, gravity
and pressure gradients induce an acoustic oscillation in the
baryon-photon fluid. Since the baryonic matter interacts
gravitationally with the dark matter, the acoustic oscillations
leaves some fingerprint in the matter power spectrum. The BAO peak
length scale is set by the sound horizon at decoupling, $\sim10^2$
Mpc. We now introduce several types of BAO data.
\\
1.~BAOI

The first one is the BAO distance measurements obtained at z = 0.2
and z = 0.35 from joint analysis of the 2dFGRS and SDSS DR7
data~\cite{Percival:2009xn}: \ba
\frac{r_s(z_d)}{D_V(0.2)}=0.1905\pm0.0061~,\\
\frac{r_s(z_d)}{D_V(0.35)}=0.1097\pm0.0036~, \ea where $r_s(z_d)$ is
the comoving sound horizon at the baryon drag epoch
$z_d$~\cite{Eisenstein:1997ik}, and \be D_V(z)=\left
[\left(\int_0^z\frac{dx}{H(x)}\right )^2\frac{z}{H(z)}\right ]^{1/3}
\ee encodes the visual distortion of a spherical object due to the
non Euclidianity of a FRW spacetime. The $\chi^2_{BAOI}$ is given by
\be \chi^2_{BAOI}=X^TV^{-1}X \ee where
\ba
X=\left(\begin{array}{c}
\frac{r_s(z_d)}{D_V(0.2)}-0.1905\\
\frac{r_s(z_d)}{D_V(0.35)}-0.1097\end{array}\right)~,
\ea
and the inverse covariance matrix
\ba
V^{-1}=\left(\begin{array}{cc}
30124.1& -17226.9 \\
-17226.9 & 86976.6 \\
\end{array}\right)~.
\ea 
With addition of the BAOI data to SN data, there is a bit
improvement of FoM, as shown in Table~\ref{II}. The best-fitted
parameters are also very close to those from the case of SN data
used only. This implies that BAOI data cannot alleviate the
degeneracy between $\Omega_{m0}$ and $w_{de}$.

2.~BAOII

Using two distance measurements of BAOI, one can derive a model
independent BAO distance ratio~\cite{Percival:2009xn} \be
\frac{D_V(0.35)}{D_V(0.2)}=1.736\pm0.065~, \ee which is independent
of $\Omega_{b0}$. For this single datapoint, one defines
\be
\chi^2_{BAOII}=\left(\frac{{D_V(0.35)}/{D_V(0.2)}-1.736}{0.065}\right)^2~.
\ee
The FoM of SN+BAOII is even smaller than that of SN data, and the
$\chi^2_{min}$ is larger than that of SN+BAOI data.

3.~BAOIII

In this case we combine the data of BAOII with the BAO distance
measurements at $z=0.275$~\cite{Percival:2009xn}: \be
\frac{r_s(z_d)}{D_V(0.275)}=0.1390\pm0.0037~. \ee These two
datapoints have  been widely-used in the literatures. In this case
the $\chi^2$ is defined as
\be
\chi^2_{BAOIII}=\left(\frac{{D_V(0.35)}/{D_V(0.2)}-1.736}{0.065}\right)^2
+\left(\frac{{r_s(z_d)}/{D_V(0.275)}-0.1390}{0.0037}\right)^2~.
\ee
As shown in Table II, the results for the combination  SN+BAOIII are
almost the same as those of SN+BAOI.

4. Distance Parameter $A$

The distance parameter $A$ is often used in the literatures, which
is the measurement of BAO peak in the distribution of SDSS luminous
red galaxies~\cite{Eisenstein:2005su}: \be
A=\Omega_m^{1/2}E(0.35)^{-1/3}\left[\frac{1}{0.35}\int_0^{0.35}\frac{dz}{E(z)}\right]^{2/3}.
\ee This quantity is  independent of $\Omega_{b0}$. The value of $A$
is determined to be $0.469(n_s/0.98)^{-0.35}\pm0.017$, where
$n_s=0.963$ is the scalar spectral index, which has been updated
from the WMAP7 data. $\chi^2$ is defined as
\be
\chi^2_{A}=\frac{(A-0.472)^2}{0.017^2}~.
\ee
The supplementary of the single datapoint $A$ to SN data gives a
large increase of FoM, because $\Omega_{m0}$ can be well constrained
and the degeneracy between $\Omega_{m0}$ and $w_{de}$ is alleviated.

5. RBAO

SDSS data can also be used to measure the radial (line-of-sight)
baryon acoustic scale, which is independent from the previous BAO
measurements which were averaged over all directions or in the
transverse direction. In that case the measured quantities are the
values of \be \Delta_z(z)=H(z)r_s(z_d)\label{dz} \ee at $z=0.24$ and
$z=0.43$~\cite{Gaztanaga:2008de}, respectively, \be
\Delta_z(0.24)=0.0407\pm0.0011\pm0.0007,~~~~\Delta_z(0.43)=0.0442\pm0.0015\pm0.0009
\ee with statistic errors and systematic errors. In this case, the
parameter $\Omega_{b0}$ is involved once again.

Fitting results show that obviously this set of data has poor
constraints on $\Omega_{m0}$ and gives poor improvement of FoM, as
shown in Table II.

%%%%%%%%%%%%%%%%%%%%%%%%%%%%%%%%%%%%%%%%%%%%%%%%%%%%%%%%%%%%%%%%%%%%%

\section{Combinations of Data Sets}
In general, to get good constraints of DE parameters one often
combines several types of observational data sets to fit
cosmological models. In this section we study the effect of
different combinations of the data sets introduced in the previous
section on constraining DE models.  We divide the combinations of
data sets into two types:
\\Type I: based on SN+BAOI+CMB, where the parameter $\Omega_{b0}$ is involved.\\
Type II: based on SN+A+R, which only depends on the distance
measurement. The calculation is much simpler than the case of Type I.

The basic combinations SN+BAOI+CMB ($B_I$) and SN+A+R ($B_{II}$) are
two widely-used combinations in the literatures. The data sets to be
combined with two basic combinations are GRB, RBAO, HUB and CBMF
data. In what follows, we  further study the properties of each data
set based on basic combinations and their effects on fitting
results; compare combinations of Type I and Type II; and find out
the combination with the largest FoM.

%%%%%%%%%%%%%%%%%%%%%%%%%%%%%%%%%%%%%%%%%%%%%%%%%%%%%%%%%%%%%%%%%%%%%%%%%%%%%%%%
\begin{table}
\begin{centering}
\begin{tabular}{|c|c|c|c|c|c|}
\hline
data & $\Omega_{m0}$ & $w_0$ & $w_1$ & $\chi^2_{min}$ & FoM\\
\hline
\hline
SN+BAOI+CMB&$0.279_{-0.010-0.023}^{+0.017+0.033}$&$-1.07_{-0.09-0.19}^{+0.14+0.27}$&$0.28_{-0.76-1.66}^{+0.46+0.84}$&544.131&27.721\\
\hline
SN+BAOI+CMB+GRB&$0.281_{-0.012-0.025}^{+0.016+0.031}$&$-1.07_{-0.08-0.19}^{+0.14+0.28}$&$0.29_{-0.78-1.74}^{+0.45+0.82}$&567.309&27.401\\
\hline
SN+BAOI+CMB+HUB&$0.275_{-0.010-0.023}^{+0.017+0.030}$&$-1.03_{-0.09-0.19}^{+0.13+0.25}$&$0.02_{-0.64-1.49}^{+0.52+0.91}$&554.034&29.495\\
\hline
SN+BAOI+CMB+RBAO&$0.274_{-0.009-0.019}^{+0.016+0.029}$&$-1.02_{-0.08-0.18}^{+0.12+0.24}$&
$0.05_{-0.71-1.52}^{+0.48+0.82}$&544.721&32.455\\
\hline SN+BAOI+CMB+CBMF
&$0.279_{-0.009-0.022}^{+0.019+0.034}$&$-1.07_{-0.09-0.20}^{+0.15+0.28}$&
$0.23_{-0.93-1.76}^{+0.51+0.90}$&587.345&26.319\\
\hline SN+BAOI+CMB+CBMF+RBAO
&$0.278_{-0.012-0.023}^{+0.013+0.025}$&$-1.03_{-0.08-0.17}^{+0.12+0.26}$&
$-0.03_{-0.70-1.57}^{+0.52+0.93}$&587.990&30.803\\
\hline
SN+BAOI+CMB+HUB+RBAO &$0.273_{-0.009-0.020}^{+0.014+0.026}$&$-1.00_{-0.09-0.18}^{+0.11+0.22}$&
$-0.08_{-0.60-1.40}^{+0.50+0.86}$&554.317&33.736\\
\hline SN+BAOI+CMB+HUB+CBMF
&$0.293_{-0.007-0.017}^{+0.013+0.025}$&$-1.04_{-0.10-0.21}^{+0.15+0.29}$&
$-0.50_{-0.85-1.71}^{+0.64+1.16}$&600.747&36.651\\
\hline SN+BAOI+CMB+HUB+RBAO+CBMF
&$0.290_{-0.008-0.017}^{+0.011+0.021}$&$-1.00_{-0.08-0.19}^{+0.16+0.28}$&
$-0.71_{-0.86-1.73}^{+0.55+1.03}$&602.267&38.815\\
\hline
\end{tabular}
\par\end{centering}
\caption{\label{tab3}The best-fitted values with 68\% and 95\% C.L. errors of $\Omega_{m0},\,w_0,\,w_1$,
and figure of merit for Type I combinations in the CPL model.} \label{III}
\end{table}

\subsection{Properties of Data Sets with Two Basic Combinations}

1.~GRB

In Table III and IV, it is obvious that the results of SN+BAOI+CMB+GRB (with GRB) and
SN+A+R+GRB are almost the same as those of SN+BAOI+CMB and SN+A+R (without GRB), respectively.
This is in accord with the results in the previous section.
We have also tested other combinations that contain the GRB data,
and arrived the same conclusion. As a result, we conclude that at
present, adding GRB data to other observational data does not
give better constraints on the DE parameters.

2. RBAO

This set of data is used only in Type I combinations.
The inclusion of RBAO data always increases the value of the best-fitted $w_0$.
In Table III,
comparing results of SN+BAOI+CMB, SN+BAOI+CMB+HUB, SN+BAOI+CMB+CBMF,
SN+BAOI+CMB+HUB+CBMF with results of those with addition of RBAO
data, we can see that adding RBAO data improves FoM well with only
small increase of $\chi^2$. Therefore it is
worthwhile to contain RBAO data in the combinations to constrain DE
models.

3. HUB and CBMF

From Table III and IV it is seen that containing HUB data can help
to improve FoM, though the increase of FoM is much less than that of
$\chi^2$ since there are 12 datapoints in the HUB data set.

For CBMF data, there are 42 datapoints. With addition of this set of
data, there is a large increase of $\chi^2$, as seen in Table III
and IV, while the FoMs for combinations with CBMF are always smaller
than those without CBMF, except the case that the HUB data is also
included in Type I combinations. As a result, it seems not necessary
to include the CBMF data to fit DE models.

But it can be seen from Table III that
based on the basic combination SN+BAOI+CMB, the inclusion of both
HUB and CBMF data can improve FoM well. For example, the FoM for
SN+BAOI+CMB+HUB+CBMF is much larger than those for SN+BAOI+CMB,
SN+BAOI+CMB+HUB and SN+BAOI+CMB+CBMF.
While the data of HUB+CBMF have no similar effect in
combinations of Type II, as shown in Table IV.
Therefore, in Type I
combinations it had better include both HUB and CBMF data, while for
Type II combinations it is not necessary to contain the CBMF data.
It is also shown that with data of HUB and CBMF the best fitted
$\Omega_{m0}$ is always larger than the ones in the cases without
HUB or CBMF.

\begin{table}
\begin{centering}
\begin{tabular}{|c|c|c|c|c|c|}
\hline
data & $\Omega_{m0}$ & $w_0$ & $w_1$ & $\chi^2_{min}$ & FoM\\
\hline
\hline
SN+A+R&$0.277_{-0.015-0.028}^{+0.014+0.029}$&$-1.01_{-0.11-0.22}^{+0.14+0.27}$&$-0.09_{-0.66-1.47}^{+0.72+1.18}$&542.642&27.930\\
\hline
SN+A+R+GRB&$0.277_{-0.014-0.027}^{+0.014+0.030}$&$-1.01_{-0.11-0.23}^{+0.13+0.27}$&$-0.05_{-0.69-1.51}^{+0.68+1.15}$&565.873&28.086\\
\hline
SN+A+R+CBMF&$0.277_{-0.014-0.027}^{+0.015+0.029}$&$-0.99_{-0.14-0.25}^{+0.12+0.26}$&
$-0.24_{-0.63-1.48}^{+0.84+1.35}$&585.775&27.181\\
\hline
SN+A+R+HUB&$0.274_{-0.013-0.025}^{+0.014+0.029}$&$-1.01_{-0.10-0.21}^{+0.14+0.27}$&$0.01_{-0.73-1.53}^{+0.56+0.99}$&550.473&30.316\\
\hline SN+A+R+HUB+CBMF
&$0.275_{-0.013-0.025}^{+0.014+0.028}$&$-1.02_{-0.09-0.21}^{+0.14+0.27}$&
$-0.01_{-0.78-1.58}^{+0.53+1.00}$&593.736&29.519\\
\hline
\end{tabular}
\par\end{centering}
\caption{\label{tab4}The best-fitted values with 68\% and 95\% C.L. errors of $\Omega_{m0},\,w_0,\,w_1$,
and figure of merit for Type II combinations in the CPL model.} \label{IV}
\end{table}

\subsection{Comparing the Two Types of Combinations}

Both basic combinations of Type I and II consist of data from Type
Ia Supernovae, Baryon Acoustic Oscillation, and Cosmic Microwave
Background Radiations. The FoM and  $\Omega_{m0}$ from two basic
combinations (SN+BAOI+CMB and SN+A+R) are very close, and
$\chi^2_{min}$ from SN+BAOI+CMB is a bit larger than that from
SN+A+R since there is a bit more datapoints in SN+BAOI+CMB. In
general, $\chi^2_{min}$ in Type II are always smaller than those of
the corresponding combinations in Type I. The fitting procedure of
Type II combinations is much simpler than that of Type I. In Type II
combinations all datapoints are only dependent on distance
measurement (i.e., one needs only to use $H(z)$). In
addition, the best-fitted $w_0$ ($\Omega_{m0}$) in Type I
combinations are always smaller (larger) than the corresponding ones
in Type II combinations.

\subsection{Recommended Combinations}

Now we take the value of FOM to be the diagnostic to quantify the
ability to constrain the DE parameters for different combinations of
observational data sets. From Table III and IV, we can see the
combinations with the largest FoM for two types of combinations are
respectively: Combination I (COM I), SN+BAOI+CMB+HUB+RBAO+CBMF for
Type I, with FoM=38.815 and Combination II (COM II), SN+A+R+HUB
for Type II, with FoM=30.316.

Though the FoM of COM II is much smaller than that of COM I, its
fitting procedure is much simpler than that of  COM I. The
combination with the best FoM in Type I is just the one that
contains the most data sets, while it is not the case in Type II
combinations. The FoM for SN+A+R+HUB+CBMF (which owes the most sets
in Type II combinations) is smaller than that for COM II (which
does not contain the CBMF data). As shown in the previous section,
adding CBMF data decreases FoM, and unlike the case in the Type I
combinations the combination of CBMF and HUB data cannot improve FoM
for Type II combinations well.

To get tight constraints of DE parameters, we therefore suggest
adopting COM I to fit DE models. On the other hand, if one prefers
to use less data, a simper procedure and quicker calculation,
COM II would be a good choice. Of course fitting results by using
COM I or COM II (or other combinations) will be a bit different.
For the CPL model, the best-fitted $\Omega_{m0}$ from COM I is a bit
larger than that from COM II.

\section{Summary}
We have analyzed the fitting properties of 11 widely-used
observational data sets and their combinations. All data sets have
been fitted with the CPL model, and the figure of merit (FoM) has
been calculated for each case to estimate constraints on $w_{de}$.
The FoM is considered as the diagnostic to quantify the ability to
constrain the DE model for different combinations of observational
data sets.

We first studied the constraint properties of 11  observational data
sets and their combinations. The main results are:

1. The case with the SN data used only gives a largish best-fitted
$\Omega_{m0}$, and  a relative small FoM,  which means that the
constraints on the DE parameters are very weak in this case.

2. The addition of HUB data to SN data can help to increase FoM and the best-fitted $\Omega_{m0}$ becomes
 much smaller than that from the case with SN data used only.

3. The addition of GRB data has almost no improvement in FoM, and
fitting results of SN+GRB have almost no difference from those of
SN. In this sense it seems not necessary to add GRB data in order to
constrain simple DE models like the CPL model.

4. The addition of 42 CBMF seems not quite useful to constrain DE
models.

5. With addition of CMB data to SN data, the FoM and constraints on
the DE parameters are greatly improved than the case with SN data
used only.

We also compare fitting results with several BAO data sets. With
these results, we studied different combinations of observational data
sets and divided them in two types, based on two widely-used basic
combinations: SN+BAOI+CMB and SN+A+R. We reach the following
conclusions:

1. It is helpful to constrain DE models by adding the RBAO data into combinations of Type I.

2. In Type I combinations the inclusion of both HUB and CBMF data
 can improve FoM well, while in Type II combinations
 including CBMF data seems not helpful.

3. The best-fitted $w_0$ ($\Omega_{m0}$) in Type I combinations are always smaller (larger) than the corresponding ones in Type II combinations.

4.  To get tight constraints of DE parameters, which means a larger
FoM, we suggest to use the SN+BAOI+CMB+HUB+RBAO+CBMF (COM I) to fit
DE models. On the other hand, if one prefers less data, simper
procedure and quicker calculation, the combination SN+A+R+HUB
(COM II) is a good choice.

Obviously COM I and COM II will lead to a bit different fitting
results of the cosmological parameters. For example, for the CPL
model the best-fitted $\Omega_{m0}$ of COM I is larger than that of
COM II. In some cases, different combinations of data sets give
rather different fitting results. This is mainly due to the fact
that at present some experiment data are not accurate enough and
sometimes there may exist tensions among them. Obtaining tight
constraints of $w_{de}$ and finding the nature of DE need more
accurate experiments.
The method of FoM used in this paper can also be applied to more new observational data\cite{Albrecht:2006um,Conley:2011ku,Sullivan:2011kv} to find good combinations of data.

\textbf{Acknowledgements}
This work was supported in part by the
National Natural Science Foundation of China (No. 10821504, No.
10975168, No.11035008 and No.11075098),  by the Ministry of Science
and Technology of China under Grant No. 2010CB833004 and by a grant
from the Chinese Academy of Sciences.

\end{document}